\documentclass[doublecol]{epl2}
\usepackage{graphicx}% Include figure files
\usepackage{amssymb}   % for math
\usepackage{dcolumn}% Align table columns on decimal point
\usepackage{bm}% bold math
\usepackage[breaklinks,colorlinks,citecolor=blue]{hyperref}
\usepackage{float}
\usepackage[mathlines]{lineno}% Enable numbering of text and display math
\newcommand{\overbar}[1]{\mkern 0.5mu\overline{\mkern-0.5mu#1\mkern-1.5mu}\mkern 1.5mu}
\newcommand{\gr}[1]{{#1}}
\newcommand{\negphantom}[1]{\ifmmode\settowidth{\dimen0}{$#1$}\else\settowidth{\dimen0}{#1}\fi\hspace*{-\dimen0}}

\title{On dispersionless transport in washboard potentials}
\shorttitle{On dispersionless transport in washboard potentials} %Insert here a short version of the title if it exceeds 70 characters
\author{I. G. Marchenko\inst{1,2} \and V. Yu. Aksenova\inst{1,2} \and I. I. Marchenko\inst{3} \and A. V. Zhiglo\inst{1}\thanks{E-mail: \email{azhiglo@uchicago.edu}}}
\shortauthor{On dispersionless transport in washboard potentials}

\institute{
  \inst{1} NSC \lq\lq Kharkov Institute of Physics and Technology\rq\rq,
 1 Akademicheskaya str., Kharkov 61108, Ukraine\\
  \inst{2} Kharkov National University,
 4 Svobody Sq., Kharkov 61077, Ukraine\\
  \inst{3} NTU \lq\lq Kharkov Polytechnic Institute\rq\rq,
 2 Kirpicheva str., 61002 Kharkov, Ukraine
}

{\gr{Fluctuation phenomena, random processes, noise, and Brownian motion}}
{\gr{Transport processes}}
{\gr{Numerical simulation; solution of equations}}

\abstract{We reassess the ``dispersionless transport regime'' of Brownian particles in tilted periodic potentials.
We show that the particles exhibit normal diffusive motion right after transitioning %from the initial locked state 
into the running state dragged by the constant bias force. No special transient dynamics appears, contrary to conjectures 
in the previous studies.
The observed flat segment in the dispersion evolution curve is solely due to the broad spatial distribution of particles
formed in the early superdiffusion stage. We quantitatively describe the whole evolution of the distribution function
during superdiffusion and the transition to the normal diffusion that follows, in the framework of the 2-well potential 
in the velocity space model.
We show that the superdiffusion exponent is $\alpha=3$. Estimate of the duration of the ostensible ``dispersionless regime''
is provided. It is shown to diverge exponentially as the temperature decreases to zero.}%in particular its exponential growth with the temperature decrease.}

\begin{document}

\maketitle

Phenomena of Brownian particle transport and diffusion in tilted periodic potentials are realized in many diverse systems. 
Superionic conductors \cite{Dieterich80Superionic}, magnetic ratchets \cite{2Tierno10PRL}, % \cy{particles moving in
optical lattices \cite{Evers13ColloidsLight},
charge-density waves \cite{Gruner88CDW}, granular gases, Josephson junctions, automatic phase-lock frequency control systems are some prominent examples \cite{3Riskenbook}.
These phenomena have been studied meticulously in recent decades \cite{Reimann02PRETiltedOverd,16LindnerSokolovBiasedUnderd,1Hanggi09RMP}.

These investigations produced a host of intriguing discoveries%results/findings/advances
, including giant diffusion \cite{4Costantini99EPL}, negative mobility \cite{5Eichhorn02PRENegMob,6Machura07PRLNegMob},
temperature abnormal diffusivity(TAD) \cite{7Marchenko12EPL,Lindenberg05NJPh}, noise-assisted transport \cite{8Borromeo05NoiseAssist}, 
stochastic resonance \cite{97MarchesoniStochResWashb}.
Dispersionless (coherent) motion of the packet of particles formed after them leaving the initial potential well under the action of the constant bias force was reported in \cite{10Lindenberg07PRL}.
We reevaluate this phenomenon in this letter.

Diffusion is quantified by the particle mean square displacement (dispersion) evolution with time. In many situations of interest this follows a power-law
$\sigma^{\prime 2}(t^\prime)\equiv \bm{\langle}|X(t^\prime)-\langle X(t^\prime)\rangle |^2\bm{\rangle}\propto t^{\prime\alpha}$.
Here angle brackets $\langle\ldots\rangle$ stand for averaging over the ensemble. $X$ is the particle radius-vector (we only consider 1D setup below) and $t^\prime$ is the time. Normal diffusion is characterized by $\alpha=1$ (Fick's law: $\sigma^{\prime 2}(t^\prime)=2D^\prime t^\prime$). Anomalous diffusion corresponds to $\alpha>1$ (superdiffusion) or $\alpha<1$ (subdiffusion)
{\cite{08VlahosAnomDif}}.%9Kumar10PREmemDifsn}}.

Anomalous diffusion mostly occurs as a transient regime; asymptotically at late times the particle ensemble spreading evolves towards normal diffusion. In such transient situations we define exponent $\alpha$ differentially as %\begin{equation*} 
$$\alpha = d\ln (\sigma^{\prime 2})/d\ln t^\prime $$ %\end{equation*} 
in what follows. Likewise, in normal diffusion regime we define the diffusion coefficient differentially as
$D^\prime=d \sigma^{\prime 2}(t^\prime)/2\,dt^\prime\,$.

We consider Langevin stochastic differential equation for $X(t^\prime)$
\begin{equation}\label{LangevinD}
  M\frac{d^2X}{dt^{\prime 2}}=-\frac{dU}{dX}
  -G\frac{dX}{dt^\prime}+F+\sqrt{2Gk_BT}\xi(t^\prime)\,.
\end{equation}
%\section{Section title }
Here $M$ is the particle mass, $G$ is the friction coefficient, $F$ is the external constant force (bias), $k_B$ is the Boltzmann constant and $T$ is the temperature.
$\xi(t^\prime)$ is the Gaussian white noise with intensity 1. %Overdot stands for time differentiation.
Potential $U(X)$ is considered periodic with period $a$, and is taken cosinusoidal $U=-U_0\cos(2\pi X/a)$ in this work. The frequency of small oscillations at the potential minima at
small $G$ is $\Omega_0=[(1/M)d^2U/dX^2|_{X=X_\mathrm{min}}]^{1/2}=(2\pi/a)(U_0/M)^{1/2}$.

In underdamped systems, characterized by $G<M\Omega_0$, two types of solutions exist for a range of $F$ in the deterministic limit $T=0$: \emph{running} solutions (which drift on average in the direction of the bias force $F$) and \emph{locked} ones (with $\langle dX/dt^\prime\rangle=0$%; $\langle\ldots\rangle|_t$ with the ``t'' subscript meaning time-averaging.)
.)
%% here a revision

When studying this (underdamped) problem in \cite{10Lindenberg07PRL}, the authors observed a horizontal plateau (superficially $\alpha=0$; cf. Fig.~\ref{EPL1disp} below) in $\sigma^2(t^\prime)$ curve, spanning $\sim$2 decades for the specified force range. The authors provided analytical arguments on the nature of that plateau, which they interpreted as a manifestation of special novel ``dispersionless transport regime''. In particular, \gr{by studying the motion of an individual Brownian particle after it leaving its initial locked state, the authors [believed to have] proved that the particle spends long time %, $\gg t_1$, \tmpnote{\{define $t_1$ before this\}}
on a trajectory that does not deviate further than $a$ from deterministic $X=V_rt^\prime+X_0$  before eventually switching to normal diffusive behavior.}% \tmpnote{+their Eq.4}

The findings of \cite{10Lindenberg07PRL} got differing interpretations in later works. ``Coherent motion'' is discussed in \cite{11SaikiaMahato09PRE} in the absence of net time-averaged particle motion ---
in the setup with the constant bias force $F$ replaced by a time-periodic force $\tilde{F}(t^\prime)$ with zero mean. 
In a series of works \cite{14SpiechowiczPRE20,SpiechowiczSciRep16,20HanggiPolFacesNonequil} the authors observe the same horizontal plateau 
in $\sigma^{\prime 2}(t^\prime)$ evolution in setups at either constant $F$ \cite{14SpiechowiczPRE20} or time-periodic $\tilde{F}(t^\prime)$ forcing. 
They call the corresponding stage of system evolution ``subdiffusion'', without checking if the known mechanisms behind subdiffusion \cite{08VlahosAnomDif} are
realized. The reason for identification of the plateau stage in $\sigma^{\prime 2}(t^\prime)$ evolution with ``subdiffusion'' stems just from the observation that the ``diffusion coefficient'' defined (\emph{in this stage, when the system does not show normal diffusive behavior}) as $D^\prime(t^\prime)=\sigma^{\prime 2}/(2t^\prime)$, decreases with time.

%---------------------------------------------

In this letter we show that in proper sense no special ``dispersionless'' physics happens in the tilted periodic potential {with Brownian particles subject to white Gaussian noise}. Nor does subdiffusion occur following the initial superdiffusion stage. Instead, Brownian particles demonstrate normal Brownian motion (normal diffusion) shortly after transitioning into the running state. The (nearly) invariance of $\sigma^{\prime 2}(t^\prime)$ and spatial distribution function $N(X,t^\prime)$
in the comoving frame are solely due to the initial broad spatial distribution formed at the superdiffusion stage%
%(physically /// its timespan... as the particles transition from locked to running state)
, with no novel physics on top. We also demonstrate how erroneous conclusions may be (and often are) drawn by visual inspection of dispersion $\sigma^{\prime 2}(t^\prime)$ curves and by improper definition of the ``diffusion coefficient'' $D^\prime$ in transient regime, in which $\sigma^{\prime 2}(t^\prime)$ deviates substantially from the normal diffusion Fick's law $\sigma^{\prime 2}=2D^\prime t^\prime$ form.

We analyze the ``dispersionless'' transport in the system governed by a dimensionless form of (\ref{LangevinD}),
\begin{equation}\label{Langevin}
 \ddot{x}=-%\frac{d}{dx}
  \sin x-\gamma\dot{x}+f+\sqrt{2\gamma Q}\zeta(t),\;
 \langle\zeta(t_i)\zeta(t_j)\rangle=\delta(t_i-t_j) \,.
\end{equation}
Overdot stands for differentiation over the dimensionless time $t=\Omega_0t^\prime$. $x=2\pi X/a$ is the dimensionless coordinate; %, $f=F/\mathrm{max}\,(dU/dX)$ is the dimensionless bias,
%{$\gamma=G/(M\Omega_0)$ is the dimensionless friction parameter.} \gr{$Q=k_BT/U_0$ is the dimensionless temperature.}
$\sigma^2$ and $D$ denote corresponding dimensionless dispersion and diffusivity. The parameters of the system are
\begin{equation}\label{dimless}
 \begin{array}{rcll}
 f&=&F/\mathrm{max}\,(dU/dX)&\left[\mathrm{bias}\right] \\
 \gamma&=&G/(M\Omega_0)&\mathrm{[friction]} \\
 Q&=&k_BT/U_0&\mathrm{[temperature]}\,.
 \end{array}
\end{equation}
The same dimensionless form is used in \cite{11SaikiaMahato09PRE,14SpiechowiczPRE20}, whereas extra factors of 2 and $2\pi$ appear in \cite{10Lindenberg07PRL}.% and \re{[Luczka]}.

For the simulations we used the friction coefficient value of $\gamma=0.03$ that is close to $\gamma_0=0.035$ used in \cite{11SaikiaMahato09PRE}, and somewhat larger than the equivalent $\gamma=\gamma_\mathrm{LSLS}/\pi\sqrt{2}\approx 0.0090$ friction coefficient to that used in \cite{10Lindenberg07PRL} (if brought to the same units as used here in Eq.~\ref{Langevin}).
The temperature parameter $Q=0.5$ is mostly used in what follows, close to equivalent $Q=2T_\mathrm{LSLS}=0.4$ of the temperature used in \cite{10Lindenberg07PRL}.
% \tmpnote{В ИСХОДНОЙ ВЕРСИИ СТАТЬИ НА [11=Spiechowicz'Revisited] ССЫЛОК В ТЕКСТЕ НЕТ].} 
$f=0.25$ for the in-depth studied setup; at this $f$ only running solutions should be realized.

Equation (\ref{Langevin}) was integrated numerically using a Verlet-type algorithm \cite{12Kuznetsov}, using the same procedure as described in \cite{13MarchenkoFconst}. The time step was selected in such a way that the maximal distance covered by the particle in one step was within $10^{-3}$ of the lattice spacing% \tmpnote{\{how can we guarantee this with the random force in RHS of (\ref{Langevin})?\}}
. The statistical averaging was performed over the ensemble of at least $5\times 10^4$ particles. The ensemble of $10^6$ particles was used to derive particle velocity and coordinate  distribution functions.

%\tmpnote{In PRE'18 we defined $D$ according to \{{\sl For each diffusivity calculation we find time $t^\prime_\mathrm{lin}$, after
%which the dispersion grows linearly with time (if averaged over the forcing
%period). The $D^\prime$ is calculated as $\sigma^2/(2t^\prime)$ at
%$t^\prime=100t^\prime_\mathrm{lin}$.\}} What was the procedure here? Like $D^\prime=d \sigma^2(t^\prime)/2\,dt^\prime$ once at the linear $\sigma^2(t)$ segment?}

Fig.~\ref{EPL1disp} shows time evolution of the dispersion of Brownian particles for different values of the external force $f$. At early times after the initial transient we see superdiffusion phase with $\sigma^2\propto t^\alpha$, $\alpha>1$. At late times normal diffusion sets in, $\sigma^2\propto t^1$. Curves 3 and 4 show the nearly horizontal section between the early superdiffusion and the late normal diffusion phase, $\sigma^2(t)\approx const$. Such sections were interpreted earlier as coherent motion phases \cite{10Lindenberg07PRL}.

\begin{figure}
\includegraphics[width=0.47\textwidth]{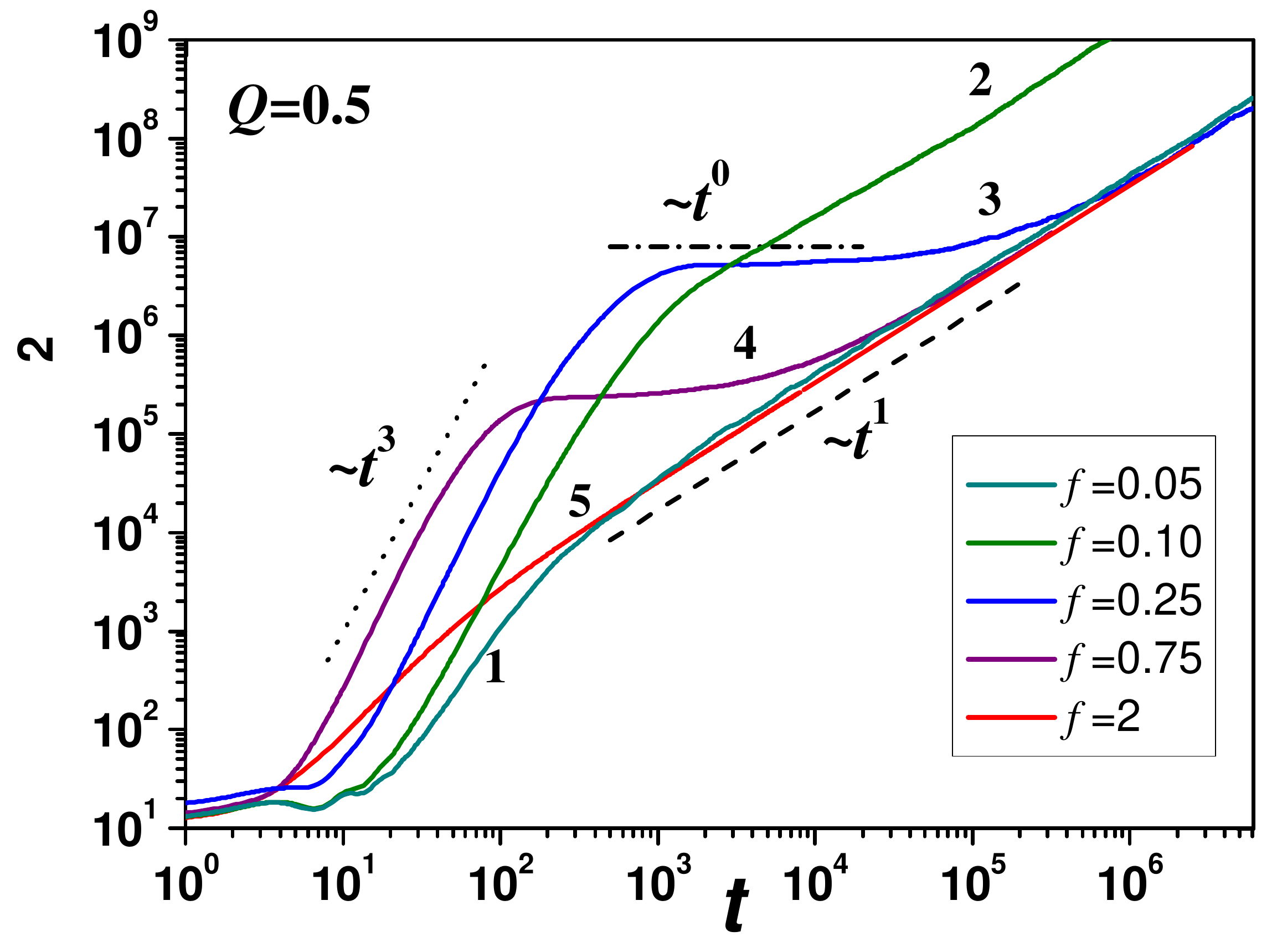}
\caption{The dispersion evolution for different values of the external force. $Q=0.5$. Power-law scaling $\sigma^2\propto t^3$ in early superdiffusion stage, $\sigma^2\approx const$ in the putative ``dispersionles regime'' and asymptotic $\sigma^2\propto t^1$ are shown at corresponding time intervals.}%\tmpnote{\{change $\sim$ into $\propto$ in Fig\}}}
\label{EPL1disp}
\end{figure}

\begin{figure}
\includegraphics[width=0.47\textwidth]{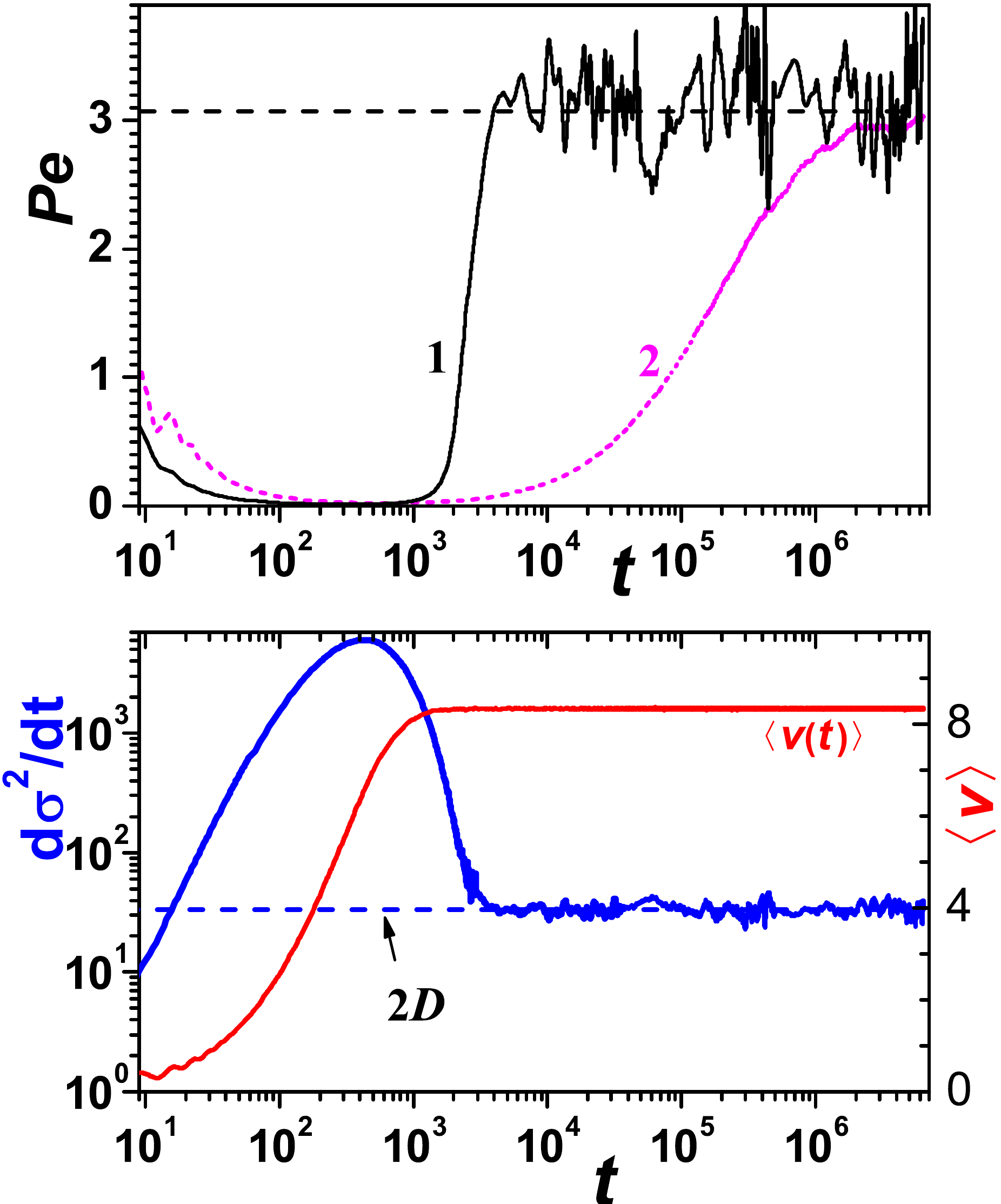}
\caption{Time dependence of the P\'eclet number $Pe(t)=2\pi\langle v(t)\rangle/D(t)$ (top), average velocity and the dispersion derivative (bottom; right and left scale, resp.). 
The horizontal dashed line in the top plot shows the asymptotic late time value of $Pe$. Curve 1 shows the $Pe$ computed with differential definition of diffusivity,
$D_\mathrm{diff}=d \sigma^{2}/2\,dt$, while curve 2 depicts $Pe_\mathrm{int}(t)$ computed with $D_\mathrm{int}=\sigma^{2}(t)/(2t)$.\\
The horizontal dashed line in the bottom plot is at the expected asymptotic value $d\sigma^2/dt|_{t\to\infty}=2D$. $f=0.25$, $Q=0.5$.}
%\tmpnote{\{change curve colors in P\'eclet subgraph\}}}
\label{EPL2PeDifIntV}
\end{figure}

%E
As the first step towards interpreting this dispersionless section, in top Fig.~\ref{EPL2PeDifIntV} we plot P\'eclet number, $Pe=\langle v\rangle l/D$, which characterizes the degree of transport coherence %\colorbox{ForestGreen}
{\cite{16Lindner01FlucNoise,02DanPRECoherentTranspPeriInhom,16Romanczuk10PRE}}. %\tmpnote{+classic}.
The higher $Pe$ corresponds to more coherent transport. Here $l$ is the characteristic spatial scale; taken the lattice spacing here, $l=2\pi$.

In the regime when the asymptotic normal diffusion has not set in, the proper way to define $D$ is via $D(t)=d\sigma^2/2\,dt$. With so defined $D(t)$ the time-dependent P\'eclet number
$Pe(t)=\langle v(t)\rangle l/D(t)$ characterizes the ratio of the particle packet average displacement over unit time to the packet diffusive broadening over the same time.
For comparison of definitions of $D$ we also show the P\'eclet number defined with integrally defined diffusion coefficient
$D_\mathrm{int}(t)= \sigma^2(t)/2t$ (as used in particular in \cite{14SpiechowiczPRE20,SpiechowiczSciRep16,20HanggiPolFacesNonequil}): curve 2 in the top Fig.~\ref{EPL2PeDifIntV}.

%G
Fig.~\ref{EPL2PeDifIntV} demonstrates that the P\'eclet number shows no spectacular features in the segment of the ``dispersionless transport''
\begin{equation}\label{t1t2}
t\in [t_1;t_2];\;\, t_1\approx 1.5\times 10^3,\; t_2\approx 3.5\times 10^4\;\mathrm{at}\;f=0.25\,.
\end{equation}
\gr{$Pe$ is nowhere infinite, as it must have been for truly coherent transport. Moreover,} the $Pe_\mathrm{int}(t)$ computed with the $D_\mathrm{int}(t)$ stays \emph{below 1} throughout the ``dispersionless regime''. Whereas the $Pe$ defined with the differential definition of $D$ (line 1) reaches the constant value of somewhat over 3 at the very beginning of the ``dispersionless'' regime, and statistically oscillates around that value into the asymptotic normal diffusion regime.

These conclusions based on the $Pe(t)$ behaviour (contradicting the reported ``dispersionless transport'' features) are corroborated by the dispersion growth rate $d\sigma^2(t)/dt$ in the $[t_1;t_2]$ interval shown in the lower part of Fig.~\ref{EPL2PeDifIntV}. We see that not only $d\sigma^2/dt$ never becomes zero, but in fact it never drops (in statistically significant manner) below the asymptotic value of $2D$.  $d\sigma^2/dt$ simply approaches that value  from above at $t\approx 4\times 10^3$, within the ``dispersionless'' interval, and statistically fluctuates around $2D$ at all later times.

\begin{figure}
\hspace{0.5em}\includegraphics[width=0.47\textwidth]{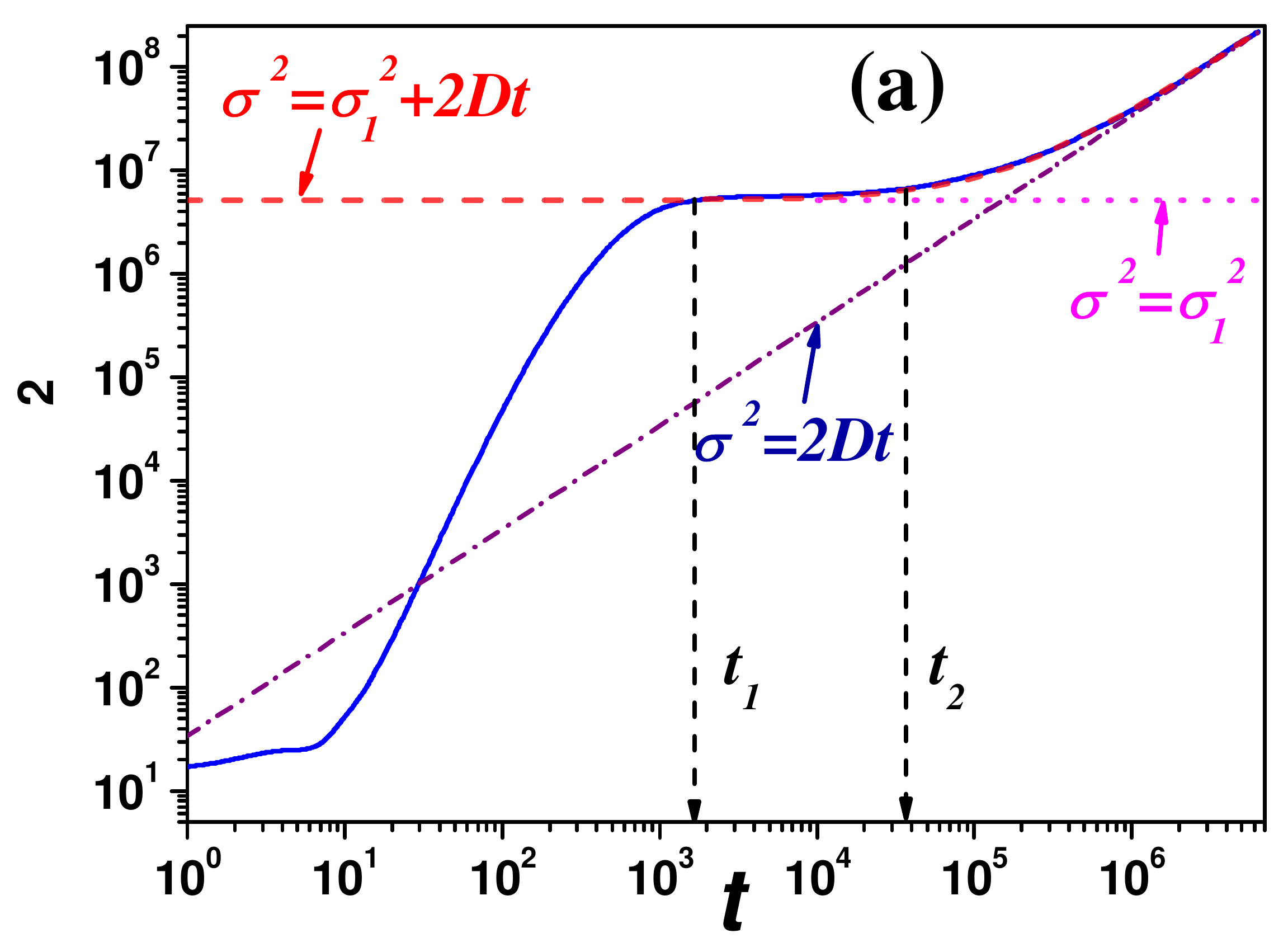}\\
\hspace{-4em}\includegraphics[width=0.48\textwidth]{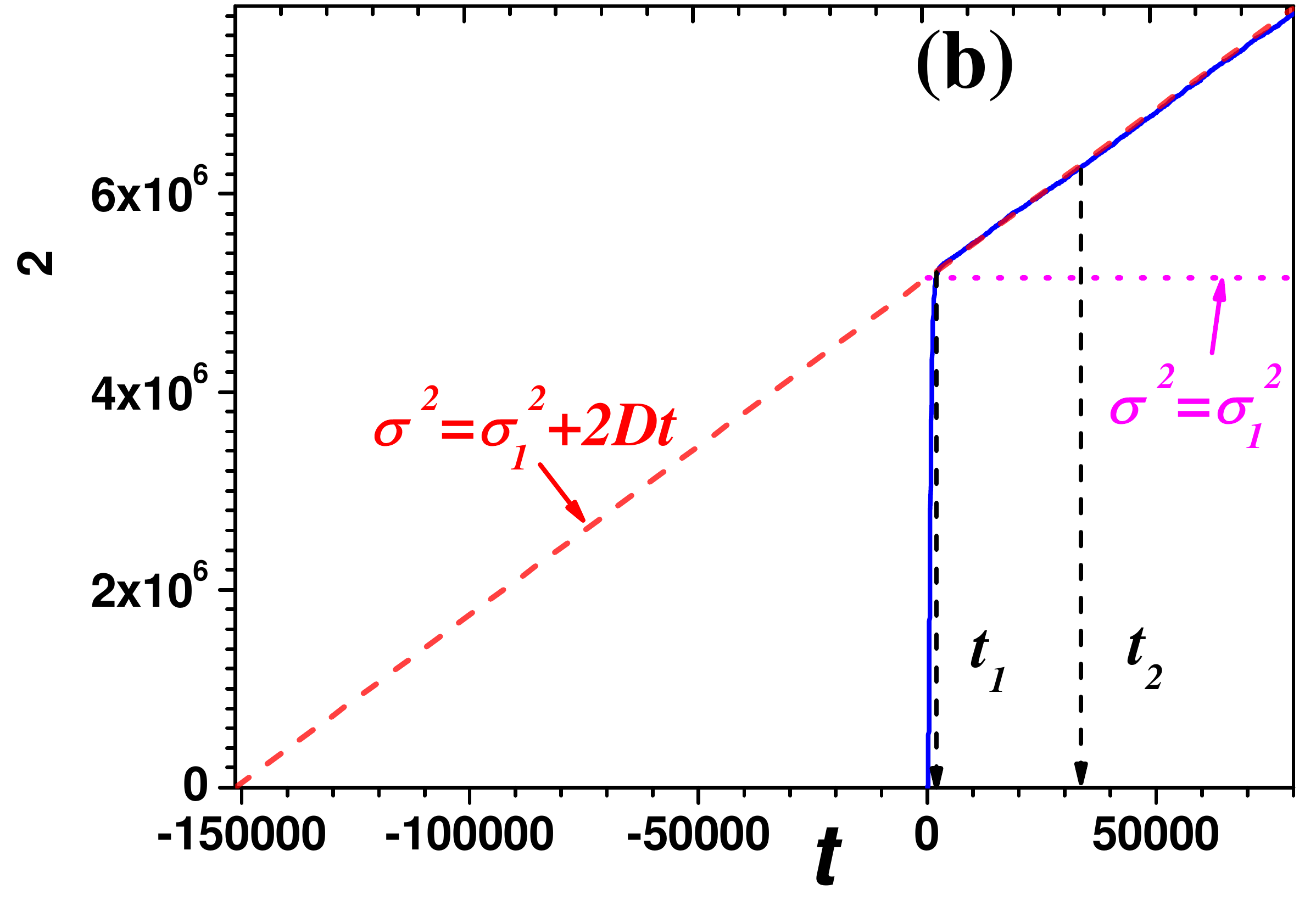}
\caption{The dispersion growth with time. $f=0.25$, $Q=0.5$.
a: log-log scale, b: linear scale.
The solid line is the simulation data. The dash-dotted line is $\sigma^2(t)=2Dt$ asymptote, the dashed line shows the approximation $\sigma^2(t)=2Dt+\sigma_1^2$, and the dotted line shows constant $\sigma_1^2$. The arrows show the starting and the ending time of the ``dispersionless'' phase, $t_1$ and $t_2$.}
\label{EPL3SigmF025}
\end{figure}

To elucidate the somewhat paradoxical look of $\sigma^2(t)$ curves in Fig.~\ref{EPL1disp} that has led different authors to searching for novel physics in time interval $[t_1;t_2]$ we compare in Fig.~\ref{EPL3SigmF025} the appearance of the dispersion evolution in log-log scale (customarily used in the references) and in linear scale. 

Superficial impression arising from Fig.~\ref{EPL3SigmF025}(a) is that $\sigma^2(t)$ undergoes rapid (superdiffusive) growth between $t\approx 10$ and $t=t_1$; the dispersion stays unchanged at $t\in[t_1;t_2]$; and eventually after $t=t_2$ normal diffusion with $\alpha=1$ ensues.
However, the flatness of $\sigma^2(t)$ curve on $t\in[t_1;t_2]$ is in a sense illusory; it is an artifact of the log-log scale and the preceding fast-growing segment of the dispersion curve. The red dashed line in upper Fig.~\ref{EPL3SigmF025} shows the linear fit $\sigma^2(t)=2Dt+\sigma_1^2$, seen to be virtually indistinguishable from the simulated $\sigma^2(t)$ through the whole ``dispersionless'' interval $[t_1;t_2]$ and the late-time normal diffusion stage.

Figure~\ref{EPL3SigmF025}(b) makes the fact of the linear $\sigma^2(t)$ growth on $[t_1;t_2]$ more overt, with linear scale used at both axes. The fast superdiffusion with $\sigma^2(t)\propto t^3$ is observed at $t< t_1$. It results in the broad distribution of particles at $t_1$ that would have required 100 times longer time to be formed if diffused according to the linear diffusion law $\sigma^2=2Dt$. With that, it takes much time, till $t_2$, for the linear growth of $\sigma^2(t)$ after $t_1$ to become apparent in the logarithmic scale used in Fig.~\ref{EPL3SigmF025}(a). Modulo this artifact of the initial broad distribution at $t_1$ --- spreading of the particle ensemble at $t\in[t_1;t_2]$ follows usual linear Fick's law, no novel physics is needed to explain the found ensemble evolution.

\begin{figure}
\includegraphics[width=0.47\textwidth]{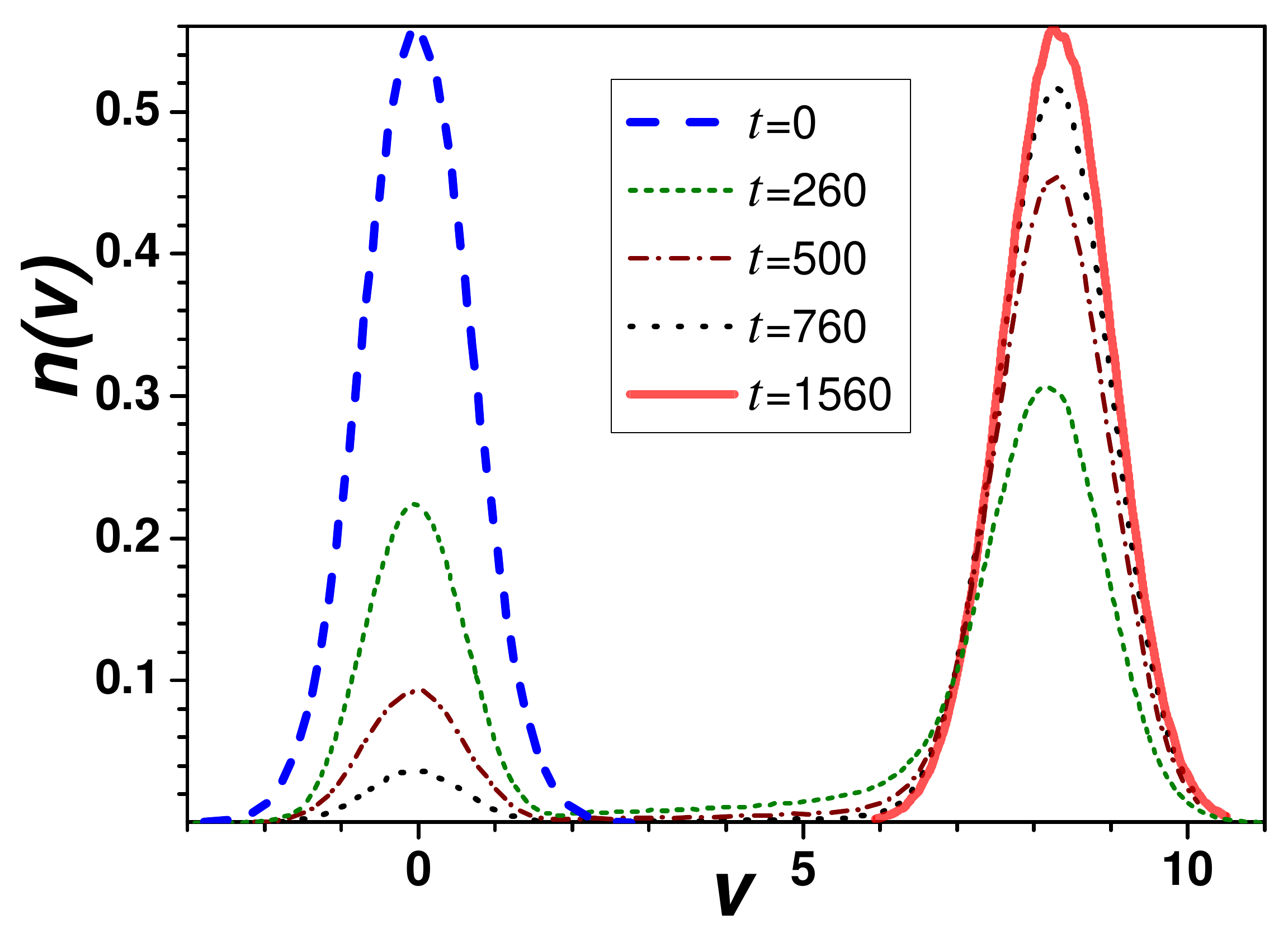}
\caption{Time evolution of the velocity distribution function. $f=0.25$, $Q=0.5$. $t=1560$ roughly corresponds to $t_1$, the start of the ``dispersionless stage''.}
\label{EPL4nvt}
\end{figure}

Let us explain the evolution of the spatial distribution function by considering the process of transitioning of the particles into the running state in the two-well velocity potential theory \cite{13MarchenkoFconst}. The effective potential $W(v)$ has two minima that correspond to locked and running particles, located at $v_l=0$ and $v_r=f/\gamma$, and a maximum between them at $v=v_{cr}(f)$ \cite{13MarchenkoFconst,15Marchenko17JETP}. In this potential, the particles transition from the locked state to the running state with the rate of $w_{lr}$ and in the opposite direction with rate $w_{rl}$.%\tmpnote{\{were called $r_-$ and $r_+$ in EPJB'14\}}

%O
Figure~\ref{EPL4nvt} shows the simulated according to Eq.~\ref{Langevin} particle velocity distribution $n(v)$  at different times. At $t=0$ all the particles are in the locked state. With time, the particles gradually transition into the running  state. This transition is virtually finished by $t=t_1$, when $n(v;t)$ assumes its final asymptotic shape, corresponding to all the particles having transitioned into the running state. This behavior (with which the ``dispersionless regime'' is observed) occurs at the parameter values at which $w_{lr}\gg w_{rl}$, so the reverse transitions from the running to the locked state can be ignored \cite{13MarchenkoFconst,15Marchenko17JETP}.

The problem is thus transformed to computation of the exit of the particles in contact with the thermal reservoir from the potential well. This is described by the Poisson process. The number of particles transitioning in $[t;t+dt]$ into the running state is $w_{lr}\exp{(-w_{lr}t)}dt$. The average velocity of such particles after the transition is $v_r=f/\gamma$. This yields exponential distribution of the running particles over $x$
\begin{equation}\label{nx2well}
n(x;t)=\frac{w_{lr}}{v_r}\exp \Bigl[ -w_{lr}\Bigl(t-\frac{x}{v_r}\Bigr)\Bigr]\theta(t)\theta(v_rt-|x|)\,,
\end{equation}
where $\theta(y)=\Bigl\{\begin{array}{ll}
1 & \mathrm{if\; }y\ge 0\\
0 & \mathrm{if\; }y<0
\end{array}$ is the Heaviside function.

\gr{In deriving (\ref{nx2well}) the two-state approximation was used, i.e. we assumed all the running particles having velocity $v=f/\gamma$, all the locked ones having $v=0$; thermal scatter of the velocities around these values neglected. This approximation is valid at sufficiently low temperatures, $Q\ll v^2_{cr}$.}

\begin{figure}
\includegraphics[width=0.47\textwidth]{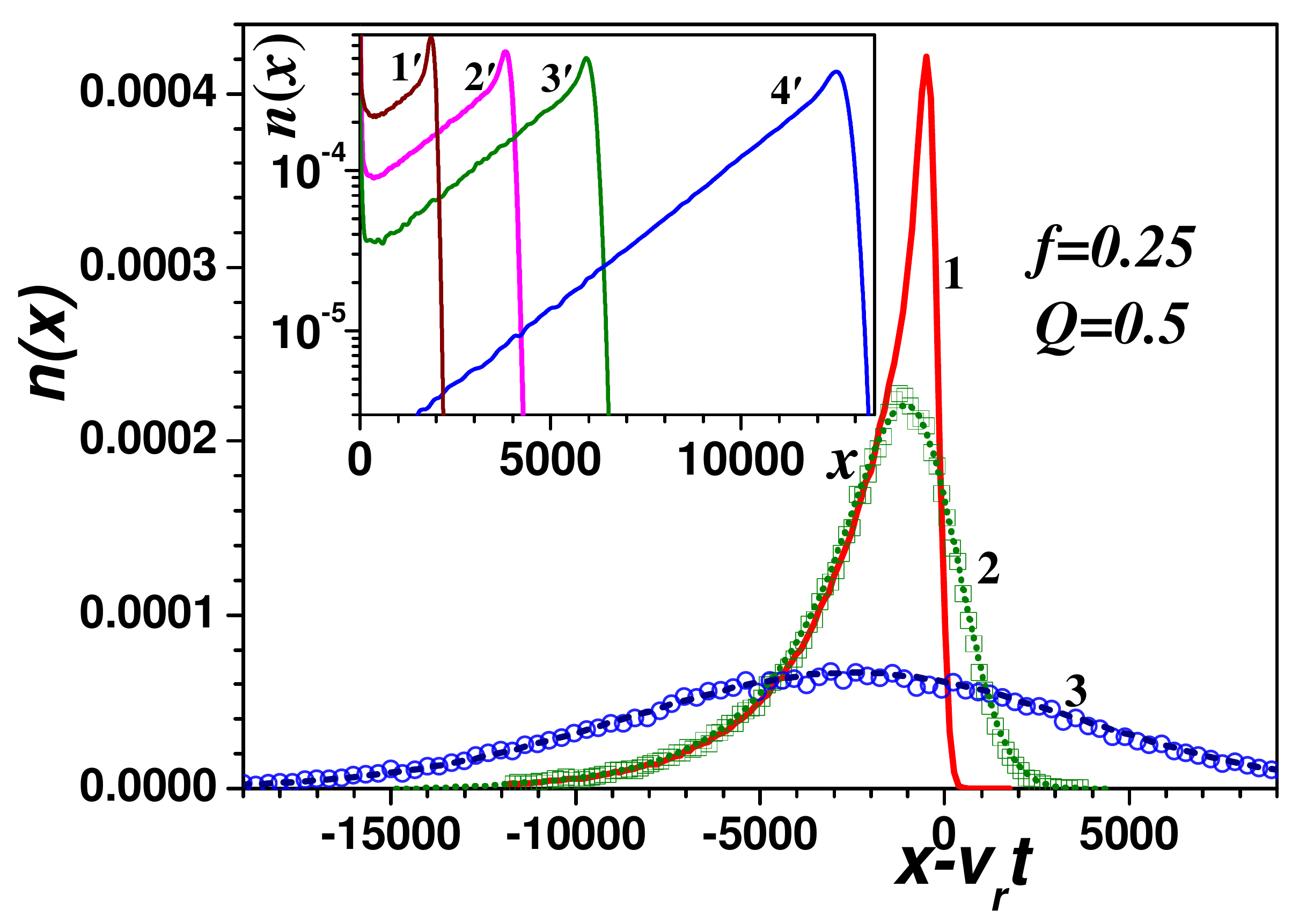}
\caption{Evolution of the spatial distribution function.\\ %$f=0.25$, $Q=0.5$.
Main figure: curve 1 and symbols (squares) 2 are simulated $n(x;t)$ in the frame moving with the velocity of the running population $v_r$ at times $t_1$ and $t_2$ respectively. Circles 3 are the simulated %%
$n(x;t)$ at $t=10^6$; by that time the profile has diffused to a close to the Gaussian shape on scales above the potential lattice constant. Dotted curves 2 and 3 were obtained by integrating the diffusion equation with constant diffusivity $D$ starting from the initial profile 1 at $t_1$ till time moments $t_2$ and $10^6$. \\
Inset: rest frame, formation of the exponential spatial $n(x)$ at early times in superdiffusion stage, when the particles gradually transition from locked to running state. Curve $4^\prime$ corresponds to time $t=1560\approx t_1$ (cf. Fig.~\ref{EPL4nvt}) when the transition is largely finished.}
\label{EPL5xtfull}
\end{figure}

The inset in Fig.~\ref{EPL5xtfull} shows the $n(x,t)$ we found in simulations. The formation of a profile of the form (\ref{nx2well}) with the steep front and exponential tail is observed. That profile mainly moves in the direction of $f$ with velocity $v_r$; its thermal broadening only becomes noticeable at significantly later times.

Given the distribution (\ref{nx2well}) its momenta may be calculated as
\begin{equation}\label{xk2well}
\langle x^k\rangle(t) =\int_0^{v_rt}{x^kn(x;t)\,dx}\,,
\end{equation}
yielding
\begin{equation}\label{sig2well}
\sigma^2(t)\equiv\langle x^2\rangle-\langle x\rangle^2= \frac{v^2_r}{w^2_{lr}}\left(1-e^{-2w_{lr}t}-w_{lr}te^{-w_{lr}t} \right).
\end{equation}

The $t\to\infty$ limit of (\ref{sig2well}) yields the dispersion at the start of the ``dispersionless regime'' (when the particles have transitioned into the running state, but the $n(x,t)$ has not yet undergone a noticeable further thermal broadening)
\begin{equation}\label{sig2wellmax}
\sigma^2_\mathrm{max} =\frac{v^2_r}{w^2_{lr}}=\frac{f^2}{\gamma^2 w^2_{lr}}\,.
\end{equation}

At early times $t \ll w_{lr}^{-1}$ Taylor expansion of (\ref{sig2well}) yields the leading power series terms
\begin{equation}\label{sig2wellearly}
\sigma^2(t)=(1/3)w_{lr}v_r^2 t^3\left[1-w_{lr}t+\mathcal{O}(w_{lr}t)^2\right].
\end{equation}
\begin{figure}
\includegraphics[width=0.49\textwidth]{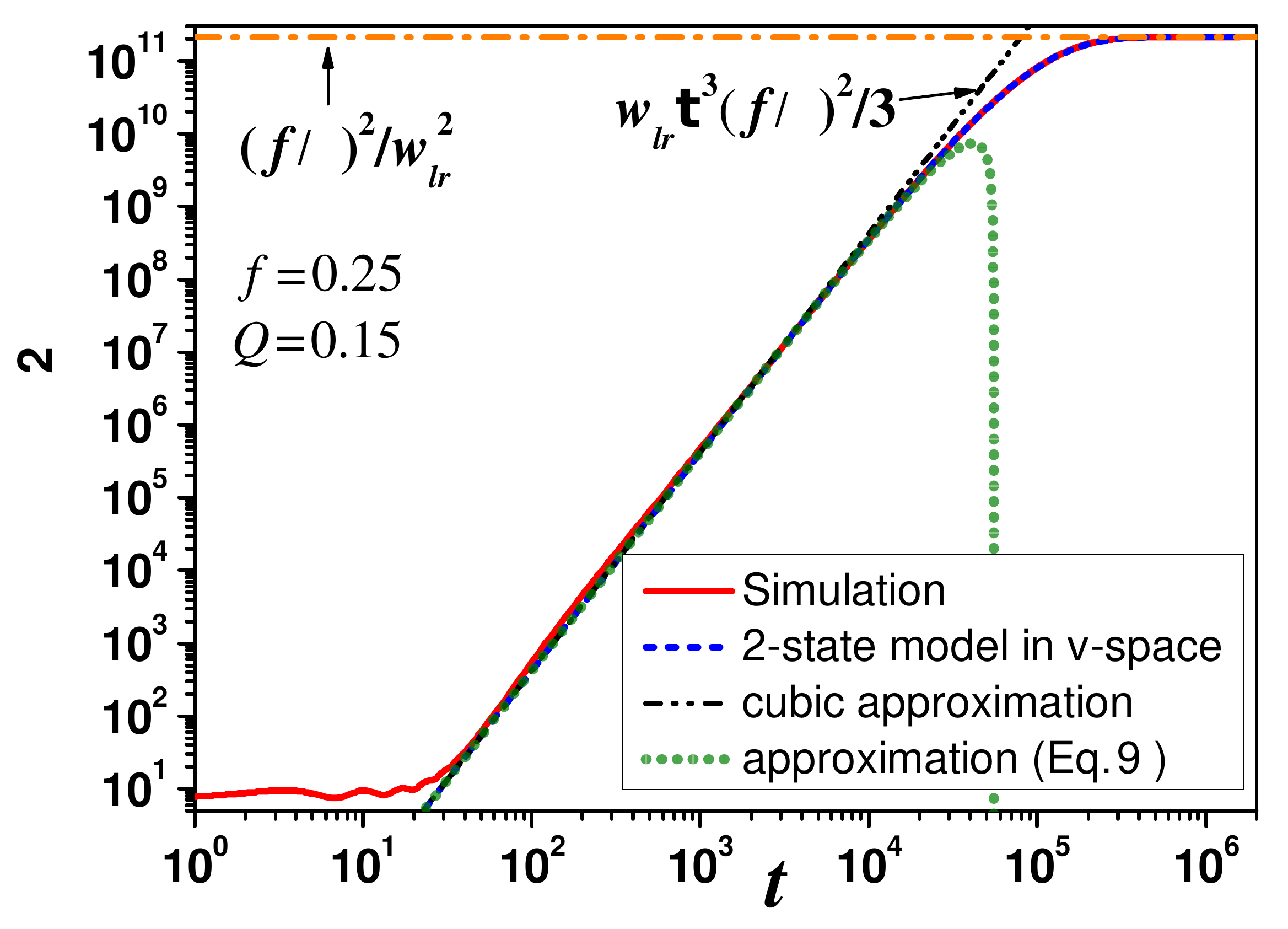}
\caption{Dispersion $\sigma^2(t)$ in superdiffusion phase. Dot-dashed horizontal line is its asymptotic value (\ref{sig2wellmax}), achieved as all particles transition into the running state. Short dashed line is the $\sigma^2(t)$ obtained in the two-state approximation in the velocity space (\ref{sig2well}). Dash-dot-dot line shows the leading cubic term in (\ref{sig2wellearly}), explaining the superdiffusion exponent $\alpha=3$ observed in Fig.~\ref{EPL1disp}, while the dotted line is the two term truncation (\ref{sig2wellearly}) that shows that noticeable deviation from cubic $\sigma^2(t)$ growth starts at $t\approx 2\times 10^4$. $f=0.25$, $Q=0.15$.}%\tmpnote{\{change Eq.No. in Fig. Smaller ``t''. Remove upper+right ticks. Righter legend. $f/\gamma$ at arrows\}}}
\label{EPL6sig2well}
\end{figure}

Predictions 
%for the $\sigma^2_\mathrm{max}$ at the ``horizontal'' ``dispersionless'' segment in the log-log representation of $\sigma^2(t)$, for the superdiffusion exponent $\alpha=3$ observed already in Fig.~\ref{EPL1disp}, and for the full quantitative $\sigma^2(t)$ growth in superdiffusion stage 
(\ref{sig2well}--\ref{sig2wellearly}) 
are demonstrated to accurately hold in Fig.~\ref{EPL6sig2well}. $Q=0.15$ was used, that yields longer superdiffusion stage than at $Q=0.5$ used in Fig.~\ref{EPL1disp}. We thus conclude that the superdiffusion phase must universally precede the ``dispersionless regime'', it must have $\alpha=3$, the whole time evolution is quantitatively understood within the 2-state model in the velocity space, with no special ``dispersionless'' physics needed to explain the simulation results.

\gr{As follows from the above}, the formation of the exponential coordinate distribution of particles is largely completed by $t_1\approx 3/w_{lr}$. After that ordinary diffusion becomes the main process slowly reshaping $n(x;t)$. The main plot in Fig.~\ref{EPL5xtfull} compares the result of the normal diffusion starting at $t_1$ with the $\sigma^2(t)$ found from simulations of (\ref{Langevin}). Solid line 1 shows the nonequilibrium distribution formed at $t=t_1$. Square and circle symbols show the numerically obtained $n(x;t)$ at $t_2$ corresponding to the completion of the ``dispersionless'' phase, and at $t=10^6$ in the late time normal diffusion regime. The dotted lines (overlapped with the corresponding sets of symbols 2 and 3) show the solution of the diffusion equation with constant diffusivity $D$ (from asymptotic late-time $\sigma^2(t)$), evolved from the initial condition given by curve 1 at $t=t_1$ to time moments $t_2$ and $10^6$. We see excellent agreement between the simulated $n(x;t)$ and the $n(x;t)$ obtained by normal diffusion of the initial profile of $t=t_1$.

In the log-log graph of $\sigma^2(t)$ the ostensible ``horizontal'' segment ends when the normal diffusive broadening of the $n(x;t)$ becomes of the order of the broad $n(x;t_1)$ width formed by the end of the superdiffusion stage. Therefore it is natural to define $t_2$ by $2Dt_2=\sigma^2(t_1)$. Accordingly, the timespan of the ``horizontal'' section is 
\begin{equation}\label{tDl}
\Delta t_{Dl}=t_2-t_1\approx t_2=\sigma^2(t_1)/(2D)=f^2/(2D\gamma^2w_{lr}^2)\,. 
\end{equation}
Since most of the particles are in the running state, $D$ is close to the particle diffusion coefficient in viscous medium, $D_\mathrm{visc}=Q/\gamma$. Hence the duration of the ``dispersionless stage'' should grow $\propto 1/(Qw_{lr}^2)$  %with the drop in temperature, 
at temperature decreasing $Q\to 0$. %, faster than $\propto 1/Q$.
According to \cite{13MarchenkoFconst,15Marchenko17JETP} 
$$ w_{lr}^2=\frac{\gamma^2v^2_{cr}}{8\pi Q}\exp{\left(-\frac{v^2_{cr}}Q\right)}\,,$$
yielding exponential divergence at low temperature $Q$
\begin{equation}\label{tDlQ0}
\Delta t_{Dl}\simeq\frac{4\pi f^2}{\gamma^3 [v_{cr}(f)]^2}\exp{\frac{[v_{cr}(f)]^2}Q} \,.
\end{equation}
The data obtained in earlier simulations %\textcolor{red}
{\cite{18Marchenko12JETPL,19Spiechowicz15PREdifAnom,MarchenkoPRE18,14SpiechowiczPRE20}} agree with the conclusion of $\Delta t_{Dl}$ growing at the temperature decreasing.

As discussed above, $w_{lr}\gg w_{rl}$ must be satisfied for the appearance of the ``horizontal'' section in the log-log graph of $\sigma^2(t)$. This imposes constraints on the form of the stationary particle velocity distribution. TAD is observed when the stationary velocity distribution has two maxima \cite{13MarchenkoFconst,15Marchenko17JETP}, both locked and running states occupied with a significant fraction of all particles. For the perceivable ``dispersionless regime'' the tilt $f$ must be above the upper limit $f_r$ of TAD-yielding tilts, so that the fraction of the particles in the locked state is negligible. As found in \cite{15Marchenko17JETP} at low temperatures $f_r=\gamma v_{cr}(1+\sqrt{2})$.  For the parameters used in this letter $f_r\approx 0.1$. Impression of ``coherent phase'' in the $\sigma^2(t)$ graph should appear at $f\in[f_r;1]$. This agrees with the behavior observed in Fig.~\ref{EPL1disp}.

%\re{$\bm{\mathbb{V}}$Conclusions} %′
In conclusion, we have analyzed the ``dispersionless transport'' \cite{10Lindenberg07PRL} of Brownian particles in the washboard potential, which received conflicting interpretations in recent works \cite{11SaikiaMahato09PRE,14SpiechowiczPRE20,SpiechowiczSciRep16,20HanggiPolFacesNonequil}. 
We showed that the whole phenomenon is due to the broad spatial distribution of the particles formed by the end $t_1$ of the superdiffusion phase, which requires significant time $t_2-t_1\gg t_1$ to be noticeably broadened further by slow normal diffusion. $t_1$ is the time needed for most of the particles to transition from the initial locked state to the running state, with $\langle v\rangle =f/\gamma$ for the running particles. The superdiffusion phase has been proven to have exponent $\alpha=3$. The evolution of the spatial ensemble dispersion $\sigma^2(t)$ during the superdiffusion phase, its transition into normal diffusion after $t_1$ have been quantitatively described in the 2-well potential in the velocity space theory \cite{13MarchenkoFconst}. We have shown that the evolution of the distribution function $n(x;t)$ follows normal diffusion equation with constant diffusion coefficient $D$ right after $t_1$; no novel transient physical behavior of particles right after transitioning into the running state (as theorized in \cite{10Lindenberg07PRL}) was \gr{observed}.

The phenomenon of ``dispersionless transport'' occurs at bias force values $f>f_r$, $f_r$ being the upper boundary of the bias at which TAD is observed. At these values in the stationary distribution in the velocity space formed after $t_1$ the vast majority of the particles are in the running state. The timespan of this ``dispersionless'' regime is given by Eq.~\ref{tDl}; it increases at the temperature $Q$ decreasing, mainly exponentially, $\propto\exp (\overbar{Q}_{f,\gamma}/Q)$.

We are grateful to Igor Goychuk for the opportunity to cross-verify our numerical
codes on test problems.

%\acknowledgments
%Insert here the text.

\bibliographystyle{eplbib}
\bibliography{epl21}
\end{document}